\documentclass{99-Styles/CCAP}
\usepackage{lineno}
\usepackage{cite}
\usepackage{textcomp}
\usepackage{times}
\usepackage{bm}         
\usepackage{amsmath,amssymb,bm,slashed} 
\usepackage{amssymb}    
\usepackage{graphicx}   
\usepackage{verbatim}   
\usepackage{color}      
\definecolor{RedViolet}{cmyk}{0.60, 0.99, 0.99, 0.0}
\definecolor{BlueViolet}{cmyk}{0.90, 0.90, 0.0, 0.0}
\definecolor{DarkBlue}{cmyk}{0.90, 0.65, 0.0, 0.0}
\definecolor{DarkGreen}{cmyk}{0.95, 0.25, 0.95, 0.25}
\definecolor{DarkYellow}{cmyk}{0.0, 0.25, 0.95, 0.25}
\usepackage{hyperref}   
\usepackage{longtable}
\usepackage{enumerate}
\usepackage{fmtcount}   

\usepackage{fancyhdr}

\fancypagestyle{firstpage}{%
  \lfoot{\small \textit{Preprint submitted to Applied Sciences}}
  \rfoot{}
  \cfoot{}
  \fancyhead{}{}
}

\begin{document}

\thispagestyle{firstpage}

\begin{tabular}{p{0.175\textwidth} p{0.5\textwidth} p{0.225\textwidth}}
    \hspace{-0.8cm}\leftline{CCAP-TN-ACCL-06}                               &
    \centering{}                                              &
    \rightline{April 13, 2021}
\end{tabular}
\vspace{-1.0cm}\\
\noindent{\color{DarkYellow} \rule{\textwidth}{0.43pt}}

\vspace{-0.125cm}
\begin{center}
  {\bf\LARGE\color{RedViolet}
    Anomalous Beam Transport Through Gabor (Plasma) Lens Prototype \\
  }
\end{center}

\makeatletter

\newcommand{\bra}[1]{\ensuremath{\langle #1 |}}   
\newcommand{\ket}[1]{\ensuremath{| #1 \rangle}}   
\newcommand{\bigbra}[1]{\ensuremath{\big\langle #1 \big|}}   
\newcommand{\bigket}[1]{\ensuremath{\big| #1 \big\rangle}}   
\newcommand{\amp}[3]{\ensuremath{\left\langle #1 \,\left|\, #2%
                     \,\right|\, #3 \right\rangle}}  
\newcommand{\sprod}[2]{\ensuremath{\left\langle #1 |%
                     #2 \right\rangle}}  
\newcommand{\ev}[1]{\ensuremath{\left\langle #1 %
                     \right\rangle}} 
\newcommand{\ds}[1]{\ensuremath{\! \frac{d^3#1}{(2\pi)^3 %
                     \sqrt{2 E_\vec{#1}}} \,}} 
\newcommand{\dst}[1]{\ensuremath{\! %
                     \frac{d^4#1}{(2\pi)^4} \,}} 
\newcommand{\tr}{\text{tr}}
\newcommand{\sgn}{\text{sgn}}
\newcommand{\diag}{\text{diag}}
\newcommand{\BR}{\text{BR}}
\newcommand\brabar{\raisebox{-2.0pt}{\scalebox{.2}{\,\,
    \textbf{(}}}\raisebox{-3.25pt}{\scalebox{.8}{{\textendash}}}\raisebox{-2.0pt}
    {\scalebox{.2}{\textbf{)}}}}
\newcommand{\gsim}      {\mbox{\raisebox{-0.4ex}{$\;\stackrel{>}{\scriptstyle \sim}\;$}}}
\newcommand{\lsim}      {\mbox{\raisebox{-0.4ex}{$\;\stackrel{<}{\scriptstyle \sim}\;$}}}

\renewcommand{\vec}[1]{{\mathbf{#1}}}
\renewcommand{\Re}{{\text{Re}}}
\renewcommand{\Im}{{\text{Im}}}
\newcommand{\iso}[2]{{\ensuremath{{}^{#2}}\ensuremath{\rm #1}}}
\newcommand{\eps}{{\ensuremath{\epsilon}}}
\newcommand{\draftnote}[1]{{\bf\color{red} \MakeUppercase{#1}}}
\newcommand{\panm}[1]{{\color{blue} #1}}
\providecommand{\abs}[1]{\lvert#1\rvert}
\providecommand{\norm}[1]{\lVert#1\rVert}

\def\parenbar{\mathpalette\p@renb@r}
\def\p@renb@r#1#2{\vbox{%
  \ifx#1\scriptscriptstyle \dimen@.7em\dimen@ii.2em\else
  \ifx#1\scriptstyle \dimen@.8em\dimen@ii.25em\else
  \dimen@1em\dimen@ii.4em\fi\fi \offinterlineskip
  \ialign{\hfill##\hfill\cr
    \vbox{\hrule width\dimen@ii}\cr
    \noalign{\vskip-.3ex}%
    \hbox to\dimen@{$\mathchar300\hfil\mathchar301$}\cr
    \noalign{\vskip-.3ex}%
    $#1#2$\cr}}}

%
\providecommand{\anmne}{\mbox{$\bar\nu_{\mu} \rightarrow \bar\nu_e$}} 
\providecommand{\nmne}{\mbox{$\nu_{\mu}\rightarrow\nu_e$}} 
\providecommand{\anm}{\mbox{$\bar\nu_\mu$}} 
\providecommand{\nm}{\mbox{$\nu_\mu$}}
\providecommand{\nue}{\mbox{$\nu_e$}} 
\providecommand{\ane}{\mbox{$\bar\nu_e$}} 
\providecommand{\enu}{\mbox{$E_\nu$}}
\providecommand{\piz}{\mbox{$\pi^0 $}}
\providecommand{\pip}{\mbox{$\pi^+$}} 
\providecommand{\pim}{\mbox{$\pi^-$}} 

\setlength{\LTcapwidth}{\linewidth}

\parindent 10pt
\pagestyle{plain}

\vspace{0.0cm}
\begin{center}
  T.~Nonnenmacher\,$^{\dag,1}$,
  T.S.~Dascalu\,$^{\dag,1}$,
  R.~Bingham\,$^{2,3}$,
  C.L.~Cheung\,$^{1}$,
  H.T.~Lau\,$^{1}$,
  K.R.~Long\,$^{4,3}$,
  J.~Pozimski\,$^{4,3}$,
  C.~Whyte\,$^{2}$,
\end{center}
\vspace{0.0cm}
\noindent\textit{\footnotesize
  \begin{tabbing}
    \hspace*{0.45cm}\= \hspace{17.5cm} \kill
     1. \> Department of Physics, Imperial College London, Exhibition Road, London SW7 2AZ, UK  \\
     2. \> Department of Physics, SUPA, University of Strathclyde, 16 Richmond Street, Glasgow G4 0NG, UK  \\
     3. \> STFC Rutherford Appleton Laboratory, Harwell Oxford, Didcot OX11 0QX, UK \\
     4. \> John Adams Institute for Accelerator Science, Imperial College London, London SW7 2AZ, UK \\
    \dag\> Corresponding author, \textit{Email:} toby.nonnenmacher14@imperial.ac.uk, t.dascalu19@imperial.ac.uk\\
  \end{tabbing}
}

\centerline{\bf Abstract}
\begin{quotation}
  \noindent

An electron plasma lens is a cost-effective, compact, strong-focusing element that can ensure efficient capture of low-energy proton and ion beams from laser-driven sources. A Gabor lens prototype was built for high electron density operation at Imperial College London. The parameters of the stable operation regime of the lens and its performance during a beam test with 1.4\,MeV protons are reported here. Narrow pencil beams were imaged on a scintillator screen 67\,cm downstream of the lens. The lens converted the pencil beams into rings that show position-dependent shape and intensity modulation that are dependent on the settings of the lens. Characterisation of the focusing effect suggests that the plasma column exhibited an off-axis rotation similar to the $m=1$ diocotron instability. The association of the instability with the cause of the rings was investigated using particle tracking simulations.

\end{quotation}

\graphicspath{ {01-Introduction/Figures} }

\section{Introduction}
\label{Sect:Intro}

One of the principal challenges that must be addressed to deliver
high-flux pulsed proton or positive-ion beams for many applications is the efficient capture
of the ions ejected from the source.
A typical source produces protons with kinetic energies of approximately
60\,keV~\cite{Peters2005,LawrieScott:2017,Faircloth_2018} and ions with kinetic energies typically below 120\,keV~\cite{Tinschert2008,Kitagawa2010}.
At this low energy the mutual repulsion of the ions causes the beam to
diverge rapidly.
Capturing a large fraction of this divergent flux therefore requires
a focusing element of short focal length.
Proton- and ion-capture systems in use today employ magnetic,
electrostatic, or radio frequency quadrupoles, or solenoid magnets to capture and
focus the beam~\cite{Chao,Nishiuchi2009,Busold2014,LawrieScott:2017}.

Laser-driven proton and ion sources are disruptive technologies that
offer enormous potential to serve future high-flux, pulsed beam
facilities~\cite{Bulanov2002,Malka2004,Daido_2012,Bin2012,pozimski_aslaninejad_2013,pozimski2016advanced,ROMANO2016,Aymar:2020drr}.
Possible applications include proton- and ion-beam production for
research, particle-beam therapy, radio-nuclide production, and ion
implantation.
Recent measurements have demonstrated the laser-driven production of
large ion fluxes at kinetic energies in excess of
10\,MeV~\cite{Zeil_2010,Prasad2011,Green2014,Dover2020}.
The further development of present technologies and the introduction of novel
techniques~\cite{Margarone2016,Morrison_2018} makes it conceivable that significantly higher ion energies will be produced in the future~\cite{Schreiber2006,Qiao2009,pozimski_aslaninejad_2013}. 
By capturing the laser-driven ions at energies two orders of magnitude
greater than those pertaining to conventional sources, it will
be possible to evade the current space-charge limit on the instantaneous
proton and ion flux that can be delivered. While in some situations the high divergence of laser-driven ion beams can be reduced~\cite{Roth2001,shmatov2003some}, for the tape-drive targets proposed for medical beams~\cite{Dover2020, Aymar:2020drr} it
necessary to capture the beam using a strong-focusing element as
close to the ion-production point as possible.

An attractive approach to providing the strong-focusing element
required to capture the low-energy ($\sim 15$\,MeV) ion flux produced
in the laser-target interaction is to exploit the strong focusing
forces that can be provided by a cloud of electrons trapped within a
cylindrical volume by crossed electric and magnetic fields.
Such an electron-plasma lens was initially proposed by Gabor in
1947~\cite{gabor1947space}.
The use of electron-plasma lenses of the Gabor type to capture and
focus proton and ion beams has been studied by a number of
authors~\cite{Mobley1979,Lefevre1979,Palkovic:1988gk,Tauschwitz1994,Goncharov2006,Goncharov2010,Schulte:2013fya}.
Such a lens has the potential to decrease the magnetic field required
in the first focusing element by a factor of more than 40 compared with
that required for a conventional beam-capture solenoid of the same focusing
strength~\cite{pozimski2005space}. Consider, for example, a 25\,MeV proton beam.  A magnetic field strength of 0.06\,T is required to achieve a focal length of 1\,m using a Gabor lens with an anode length of 0.3\,m. To achieve the same focal length using a solenoid requires a field of 2.6\,T. Such strong focusing is particularly important when beams are produced with a large divergence angle~\cite{pozimski2016advanced}.
The Gabor lens is therefore the ideal focusing element by which to
capture a laser-accelerated proton or ion beam.
Its compactness and relatively low price are key if it is to be
exploited in particle-beam therapy facilities. Furthermore, it has been shown in simulation that a Gabor-lens-based system is capable of capturing laser-generated proton beams at energies as high
as 250\,MeV, the energy required to serve a proton-beam therapy
facility~\cite{pozimski_aslaninejad_2013}.

Following the initial proposal by Gabor, several groups have reported
stable operation of a space-charge lens under a variety of electrode
and magnetic field 
configurations~\cite{Malmberg1975,Mobley1979,Lefevre1979,Malmberg1980,Palkovic:1988gk}.
Experiments with ion beams confirmed the focusing capability of the
Gabor lens and observed emittance
growth~\cite{Palkovic:1988gk,Pozimski1992,Schulte:2013fya}.
The mechanism for electron production and the
inhomogeneity in the electron density within the lens were believed to
cause the observed growth in emittance~\cite{Pozimski1992}.

The focusing strength of a Gabor lens is determined by the electron
density.
The theoretical maximum electron density is related to the electric
and magnetic field strength~\cite{pozimski_aslaninejad_2013}.
Careful design of the field configuration allowed certain lenses to
operate at electron densities of 61\% of the theoretical
maximum~\cite{pozimski2005space}. At high pressure, the electrons are lost due to the radial expansion of the plasma driven by collisions with neutral atoms. At low pressure, this radial transport can be caused by small azimuthal asymmetries in the applied electric or magnetic fields~\cite{Malmberg1980}. Further work was directed towards the design of an electrostatic lens for
the space-charge-compensated transport of a high intensity heavy ion
beam~\cite{Goncharov2005}.
In this case, the absence of emittance growth due to the lens was
reported~\cite{Goncharov2005}.

Advances in simulation and finite-element analysis have been
exploited to calculate the expected focusing strength of a
space-charge lens and to study the resulting phase-space
transformation on a beam passing through the lens.
Good agreement between experimental results and beam-transport
simulations~\cite{pozimski2005space,Schulte:2013fya} suggest
that plasma instabilities are a likely cause of beam aberrations.
Experimental observations and numerical results~\cite{Meusel:2013}
have confirmed that the confined plasma is vulnerable to the diocotron
instability~\cite{Levy1965}.
Further studies are required to characterise the configurations under
which a Gabor space-charge lens operates in a stable regime.

Plasma-lens focusing for electron beams is being developed by the CERN
Linear Electron Accelerator for Research (CLEAR)
collaboration~\cite{sjobak2019status}.
Evidence of aberrations in the CLEAR lens due to radial non-uniformity
of the plasma temperature has been observed~\cite{lindstrom2018emittance,lindstrom2018overview}.  
Previous work at the Stanford Linear Accelerator Center (SLAC)
demonstrated the plasma-lens focusing of 28.5\,GeV electron~\cite{joshi2002high} and positron~\cite{Hogan2003} beams. 
Numerical simulations were able to describe the observed non-linear
focusing force in this experiment~\cite{muggli2008halo}. Discharge-capillary active plasma lenses were also investigated as compact devices for focusing 100 MeV-level electron beams produced by a gas-jet-based laser-plasma accelerator. Both weak and strong chromatic effects~\cite{LBNL_APL2015} were observed with the potential to cause emittance degradation~\cite{LBNL_APL2017}.
Research at SLAC seeks to demonstrate the use of such a plasma lens for staging in plasma
wakefield acceleration or in radially-symmetric final focusing for
linear colliders~\cite{doss2019laser}. As compact and tunable devices, active plasma lenses~\cite{Barov1998,Pompili2017} are a promising solution for the extraction and transport of the witness bunch while removing the driver without loss of beam quality~\cite{Chiadroni2018,Pompili2019,Dotto2020}. Finally, electron plasmas were studied in multicell traps~\cite{Danielson2006, Hurst2019} to develop methods by which the number of accumulated positrons could be increased compared to the present limits coming from the requirement for long-term confinement.

In this paper we report the performance of a Gabor lens prototype
constructed at Imperial College London.
The lens was exposed to 1.4\,MeV protons at the Surrey Proton Beam
Centre~\cite{SurreyIonBeamCentre}.
The effect on the beam is presented and compared to the results
of a simulation of the impact of plasma instabilities on the focusing
forces produced by the lens.

\graphicspath{ {02-Gabor-Lens/Figures/} }

\section{The Gabor Lens} \label{The Gabor Lens}

\begin{figure}
  \begin{center}
    \includegraphics[width=0.9\textwidth]{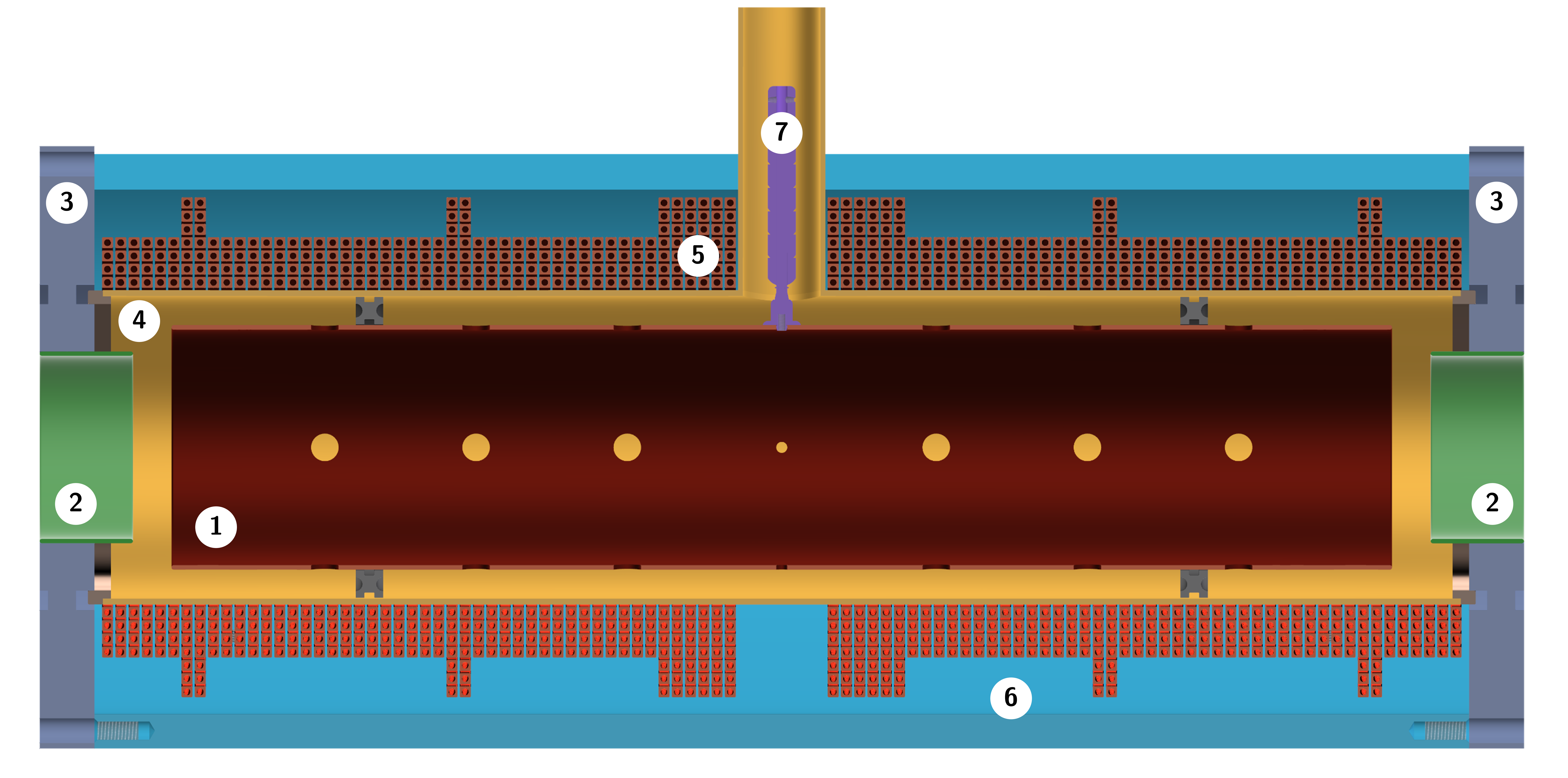}
  \end{center}
  \caption{
    Internal structure of the IC Gabor lens viewed in longitudinal cross-section. The main components are: 1-central anode, 2-end electrodes, 3-end flanges, 4-vacuum tube, 5-pancake coils, 6-outer tube, 7-high-voltage feed-through.
  }
  \label{ICGaborLens}
\end{figure}

A schematic of the prototype Gabor lens is shown in figure~\ref{ICGaborLens}.
The total length of the lens, from end flange to end flange, was
540\,mm.
The central anode was formed of a copper cylinder with an inner
diameter of 85.7\,mm and a length of 444\,mm.
The copper cylinder was 1.6\,mm thick and had four rows of
four 10\,mm diameter holes forming lines along the axis
of the cylinder spaced by $90^\circ$ so that the volume inside the
anode would be evacuated efficiently. Two ceramic isolating spacers were used to maintain the position of the central high-voltage electrode and to electrically isolate it from the vacuum tube. A 15\,mm copper high-voltage (HV) connector was soldered to the central electrode to provide a socket for the high-voltage feed-through designed for voltages up to 60--65\,kV.

The two end electrodes were formed of two copper cylinders with a length of 34\,mm, an inner diameter of 66.7\,mm, and a wall thickness of 1.6\,mm. The ends of the cylinders were manufactured with rounded edges to reduce the likelihood of sparking, with a minimum gap of 16.8\,mm to the high-voltage anode. The end electrodes were press fitted into the mild steed end flanges. The outer tube and end flanges were connected to ground.

The configuration of the pancake coils is shown in figure~\ref{ICGaborLens}. The input and output tails of the coils exited through the 50\,mm gap in the outer tube of the lens. A water cooled copper conductor was used with a square cross-sectional area of 10.87\,mm$^2$. The base configuration of the coil included four windings.  The number of windings was locally increased to seven at specific positions to generate a more uniform magnetic field. A maximum magnetic field of approximately 55\,mT was achieved at 45\,A.

A power supply (Glassman LP 60--46) was used to regulate the current that flows through the coils. Typical values for the current were in the range of 14\,A to 30\,A. The voltage for the central electrode was provided by a high-voltage supply of the Glassman Series FR type with typical values of between 8\,kV and 20\,kV.  

The pumping system was comprised of a roughing pump (Edwards 5) and a turbo molecular pump (Leybold Turbovac 151) with a pressure gauge (Leybold Penninvac PTR 90 N) and a pressure-gauge monitor (Leybold Graphix One). The lowest pressure achieved was $3 \times 10^{-7}$\,mbar, with pressures of up to $3 \times 10^{-5}$\,mbar when a non-neutral plasma was established inside the lens. Settings of HV and current that produced a stable plasma could be distinguished clearly from those which gave rise to an unstable plasma at low pressure ($\sim 10^{-7}$\,mbar). For higher pressures, this distinction was more difficult to observe. 

Plasma in the lens was produced by increasing the high voltage applied to the anode and the current in the magnetic coils. A significant increase in pressure was observed when a stable plasma was first established in the lens. Simulation of the plasma discharge within the lens indicated that a high electron density, $\sim 5\times10^{-7}$\,Cm$^{-3}$, was produced.

\graphicspath{ {03-Initial-studies/Figures/} }

\section{Plasma Characterisation}
\label{Sect:Plasma_char}

\begin{figure}
  \begin{center}
    \includegraphics[width=0.75\textwidth]{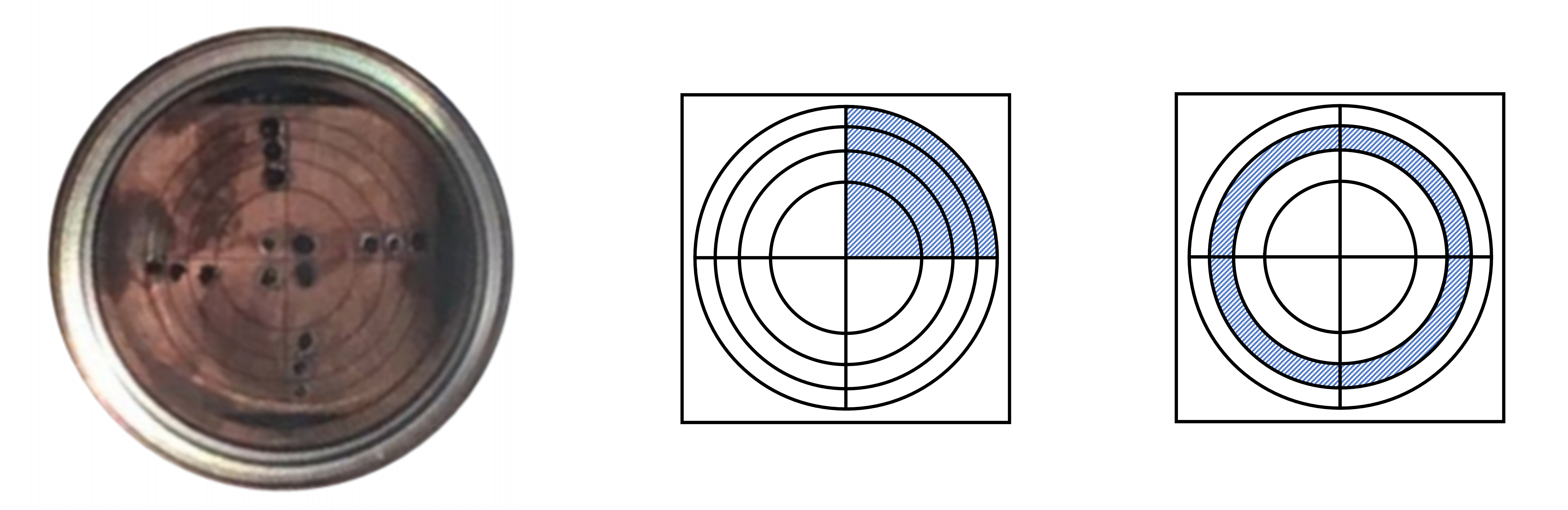}
  \end{center}
  \caption{
    The segmented detector used for measuring the current of electrons and ions exiting the Gabor Lens. The detector is divided in to 16 sections of equal area which were combined in sector (middle) or concentric circle (right) arrangements. 
  }
  \label{Medusa}
\end{figure}

\begin{figure}
  \begin{center}
    \includegraphics[width=0.9\textwidth]{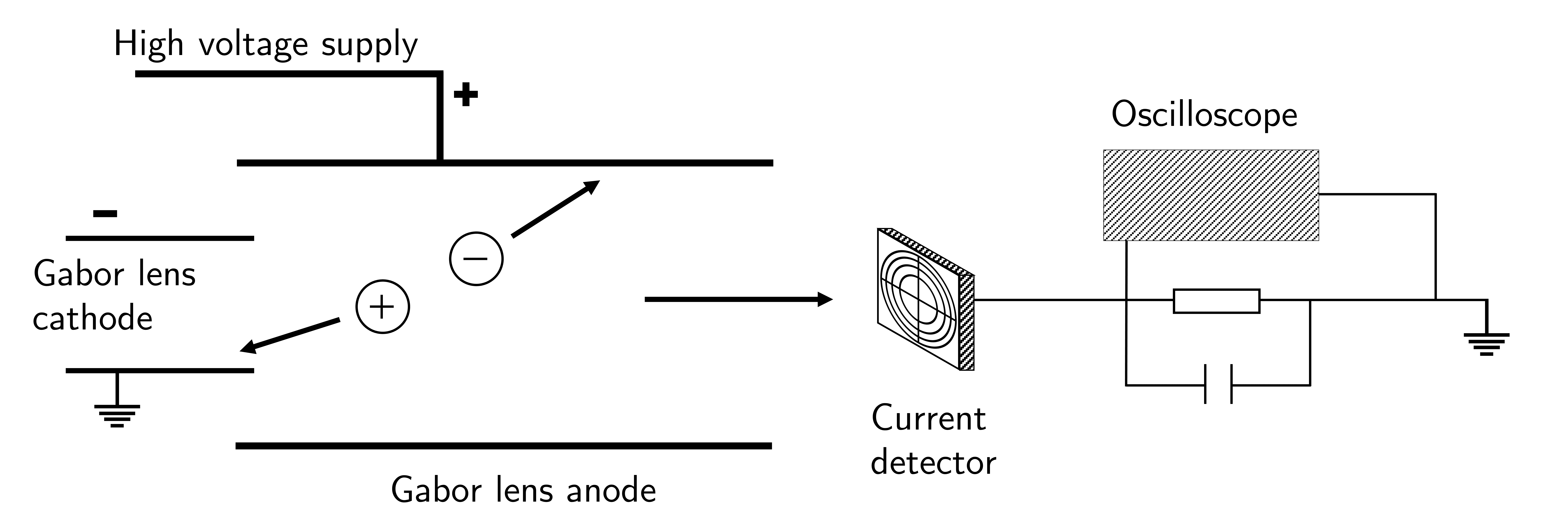}
  \end{center}
  \caption{
    Schematic of the Gabor lens, current detector, and oscilloscope.
    The high voltage supply maintains the drop across the electrodes.
    Expelled ions hit the detector, and the current signal is
    converted to a voltage output signal in the oscilloscope.
  }
  \label{fig:Schematic}
\end{figure}

The operation of the lens was tested over the range of available anode voltages and coil currents to identify the regime for which a stable plasma could be produced. Measurements of the plasma in the lens were made using the Medusa voltage sensor shown in figure~\ref{Medusa}. The sensor detects the current of ions and electrons discharged by the lens and was composed of 16 equal segments with a total area of 122.5\,mm$^{2}$. The detector segments were connected either in concentric circles or in a sector arrangement, with four segments combined and fed into one channel of an oscilloscope.
The Medusa detector was used to characterise the range of high voltage and current settings which would produce a stable plasma within the lens.
A schematic diagram of the experimental setup for these primary studies is given in figure~\ref{fig:Schematic}. For a constant current through the coils of the lens, producing a constant magnetic field, the high voltage was increased from zero until plasma was produced in the lens. The presence of a stable plasma in the lens was indicated by a steady voltage read from the Medusa detector. The high voltage was then increased further, until instability in the plasma, characterised by sparking, was observed as an extreme variation in the output voltage reading.

\begin{figure}
  \begin{center}
    \includegraphics[width=0.55\textwidth]{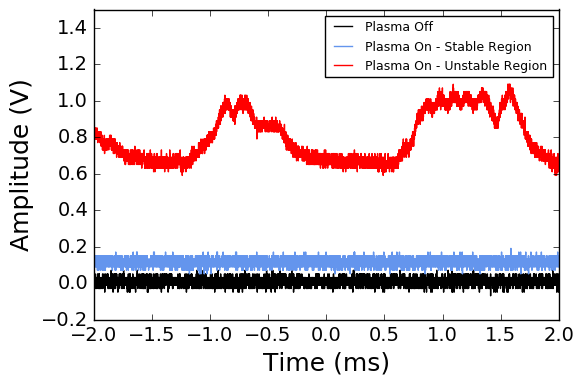}
  \end{center}
  \caption{
    Amplitude of signal from the Medusa detector in three
    regions: plasma off, plasma on, and plasma on in unstable region.
    The time range is given 0.0004\,s, and the time resolution of the
    measurement is $4 \times 10^{-7}$\,s.
    The voltage range is 0.5\,V with a resolution of 0.01\,V.}
  \label{fig:Medusa_Amplitude}
  \label{fig:Medusa_Fourier}
\end{figure}

Figure~\ref{fig:Medusa_Amplitude} shows the amplitude responses observed in the Medusa detector that are typical of three modes of operation:
\begin{itemize}
  \item\textit{Plasma off}: high voltage and current through coils
    below the threshold for plasma to be produced; 
  \item \textit{Stable plasma}: plasma produced with high voltage
    below $25$\,kV and current below $27$\,A; and
    \item \textit{Unstable plasma}: plasma produced with higher
      magnetic field causing considerable sparking and therefore large variations in the output amplitude.
\end{itemize}

The mean and standard deviation of the Medusa voltage measurement of the different plasma regions are
shown in table~\ref{TableValues}.
The mean increases slightly when the plasma is switched on, while the standard deviation remains largely unchanged.
As the current applied through the coils becomes large, and the unstable region is reached, the mean and standard deviation rose and a large amount of noise was observed.
This is shown in figure~\ref{fig:Medusa_Fourier}, where the level of noise is similar with the plasma off and the plasma on. The noise increases appreciably only when the unstable region is reached. The frequency spectrum of the voltage signal was studied in each of the three modes of operation. In the unstable region, the low-frequency noise is increased significantly while the high-frequency noise remains largely unchanged. The characterisation of the operating regimes described in this section was used to verify that the lens produced a stable plasma during the beam test.

\begin{table}
  \caption{
    Mean and Standard Deviation values for Plasma Off, On, and
    Unstable regions shown in figure \ref{fig:Medusa_Amplitude}.
  }
  \label{TableValues}
  \begin{center}
    \begin{tabular}{|l|l|l|}
      \hline
          & \textbf{Mean (V)} & \textbf{\begin{tabular}[c]{@{}l@{}}Standard \\ Deviation (V)\end{tabular}} \\ \hline
      \textbf{Plasma Off}      & 0.008 & 0.019
      \\ \hline
      \textbf{Plasma On}       & 0.114 & 0.020
      \\ \hline
      \textbf{Plasma Unstable} & 0.797 & 0.133
      \\ \hline
    \end{tabular}
  \end{center}  
\end{table}

\graphicspath{ {04-Beam-test-setup/Figures/} }

\section{Beam Test Setup}
\label{beamtestsection}

\begin{figure}
  \begin{center}
    \includegraphics[width=0.8\textwidth]{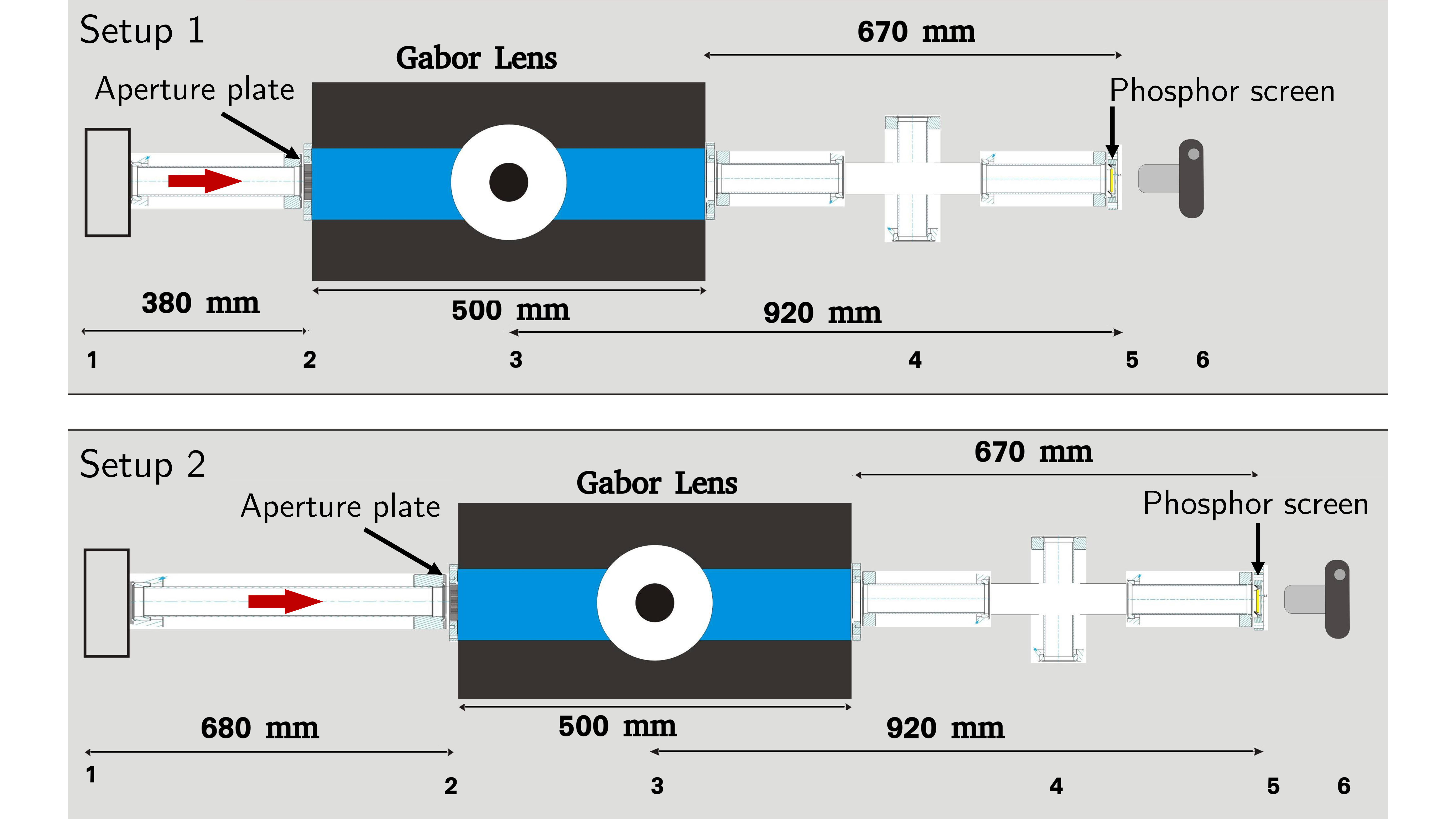}
  \end{center}
  \caption{
    Schematics of the day 1 beam test setup, Setup 1 (top) and the day
    2 beam test setup, Setup 2 (bottom).
    The setup includes the Gabor lens, aperture, and beam pipes.
  }
  \label{fig:beamtest}
\end{figure}

The prototype Gabor lens was exposed to proton beams with a kinetic
energy of 1.4\,MeV at the Ion Beam Facility at the University of
Surrey~\cite{SurreyIonBeamCentre} in October 2017.
Schematic diagrams of the two setups used in the beam tests are shown
in figure~\ref{fig:beamtest}.
The proton beam entered the lens through a section of evacuated beam
pipe.
The length of the drift on the first day of data taking was
approximately 380\,mm (Setup 1 in figure \ref{fig:beamtest}).
On the second day the length of the drift was extended to
approximately 680\,mm to exploit the divergence of the beam to
illuminate a larger area at the front face of the lens (Setup 2 in figure \ref{fig:beamtest}).

\begin{figure}
  \begin{center}
    \includegraphics[width=0.4\textwidth]{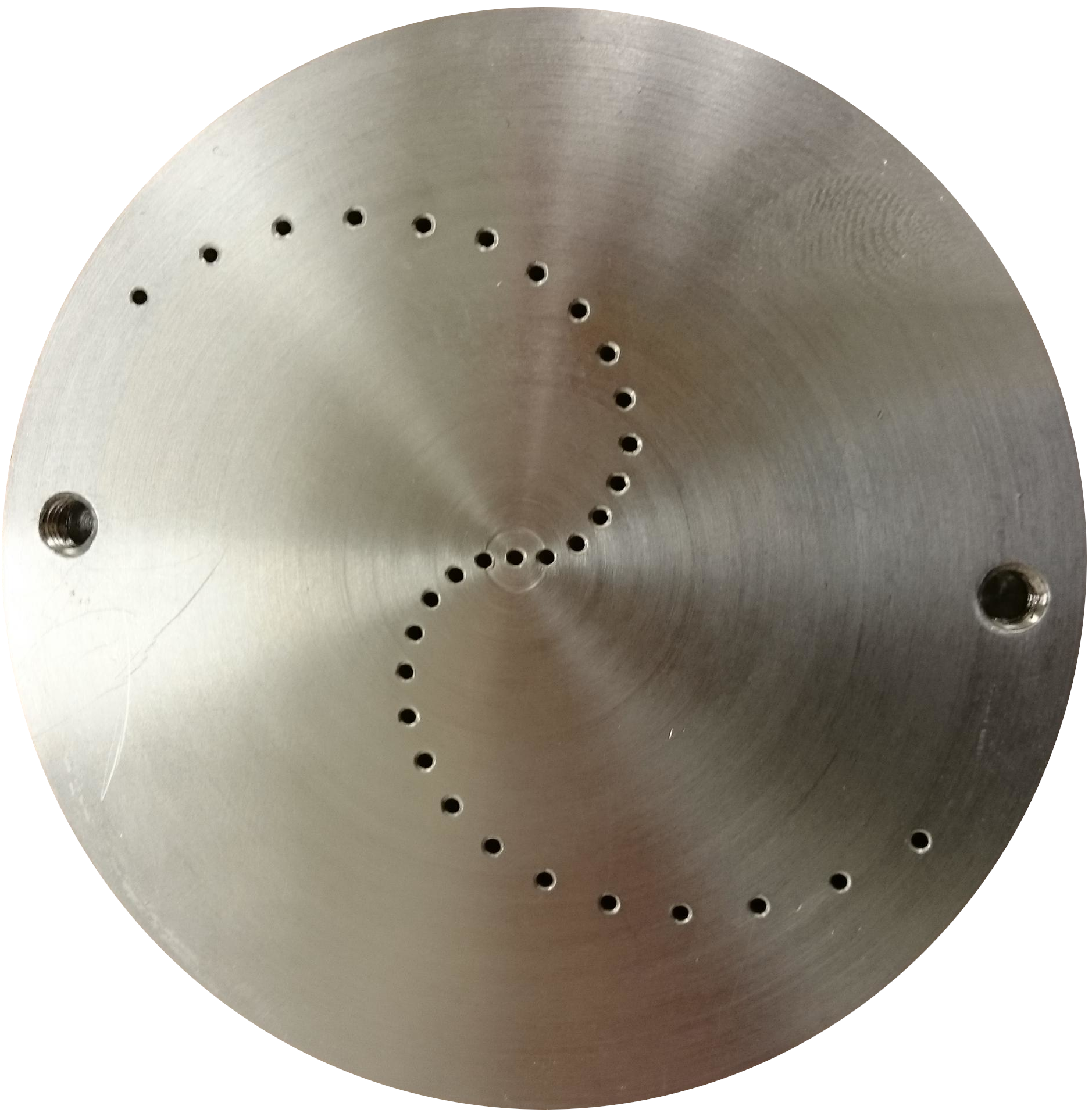}
  \end{center}
  \caption{
    Photograph of the aperture placed in the beamline upstream of the
    Gabor Lens.
    30 holes of 2~mm width are drilled in a symmetrical pattern around
    one further hole on the axis.
    The surrounding holes are pitched at an angle of 20$^\circ$.
  }
  \label{fig:aperture}
\end{figure}

Narrow ``beamlets'' were created using an aperture plate placed at the
entrance to the lens (see figure~\ref{fig:aperture}).
The holes in the aperture plate were 2\,mm in diameter
and arranged in a pattern designed to minimise the overlap of the
outgoing beamlets under a focusing force that is rotationally
symmetric about the beam axis. 

A further section of evacuated beam pipe of length 670\,mm
was attached to the downstream flange of the prototype lens.
A phosphor screen was installed on the downstream flange as indicated
in figure~\ref{fig:beamtest}.
The phosphor screen used was a P43 phosphor surface on an aluminised
pyrex substrate with an effective area of diameter 44.9\,mm and a
thickness of 10-15\,$\mu$m.
Photographs of the image of the beam on the phosphor screen were
acquired with a DSLR camera using an exposure long compared to the
beam spill.
\begin{figure}
  \begin{center}
    \includegraphics[width=0.8\textwidth]{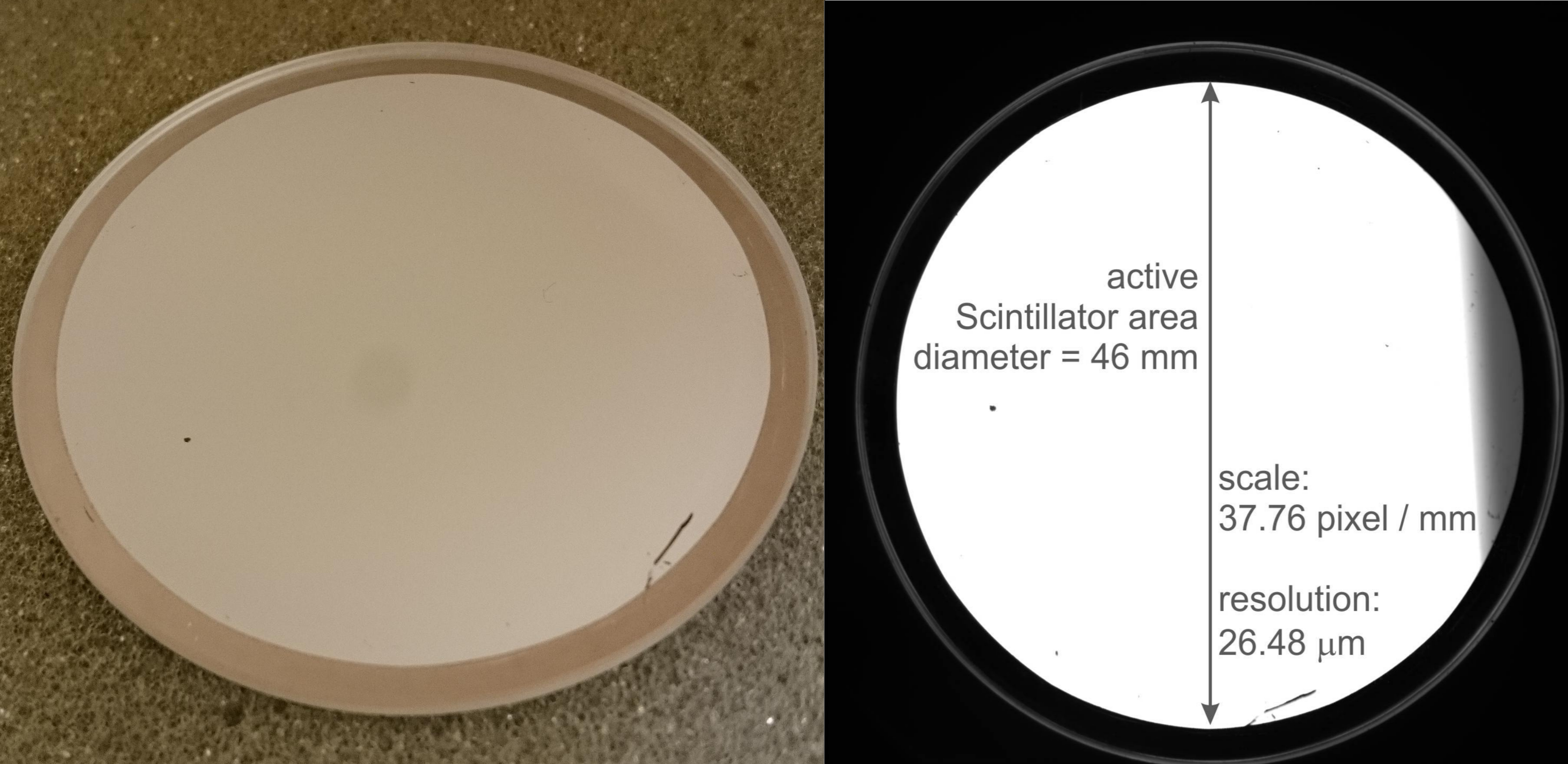}
  \end{center}
  \caption{
    Photograph and schematic of the phosphor screen used for imaging
    the beam.
    The screen was composed of a P43 phosphor surface on a substrate
    of aluminized pyrex, and the scale and resolution of the screen
    are shown on the schematic.
  }
  \label{fig:scintillator}
\end{figure}

\graphicspath{ {05-Analysis/Figures/} }
\section{Characterisation of lens performance}
\label{Sect:LensChar}

\begin{figure}
  \begin{center}
    \includegraphics[width=0.8\textwidth]{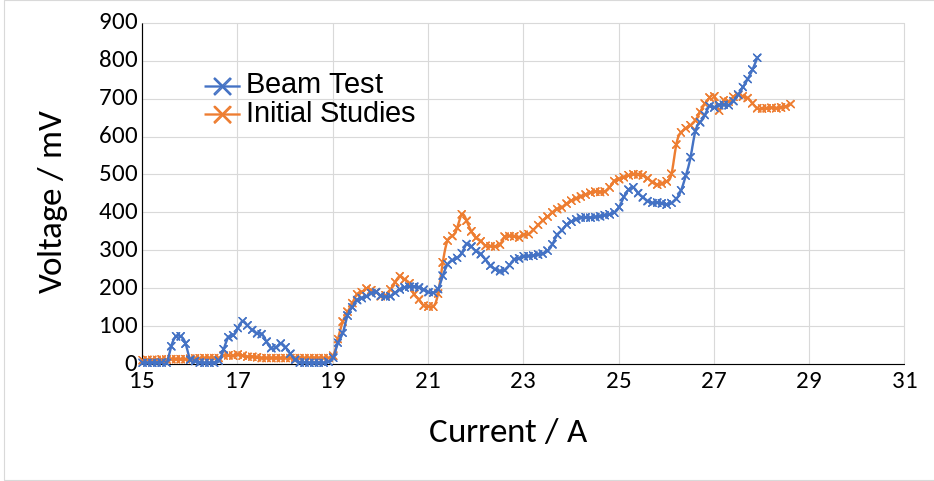}
  \end{center}
  \caption{
    Voltage-current characteristic of the Gabor lens measured with the
    lens on the beam line (blue crosses) compared to measurements in
    the laboratory at Imperial (orange crosses).
    During the beam test the gas pressure in the lens was 10\%
    higher than the pressure at which the lens operated in the
    laboratory.
  }
  \label{fig:SurreyLabComparison}
\end{figure}

The lens was set up on the beam line and stable operation of the lens was
established as described in section~\ref{Sect:Plasma_char}.
The voltage-current characteristics of the lens measured using the
Medusa detector with the lens on the beam line are compared with those
measured at Imperial in figure~\ref{fig:SurreyLabComparison}.
The two sets of measurements show similar features indicating that the lens was
operating in a similar manner to the operation in the lab at Imperial. Figure~\ref{fig:SurreyLabComparison} was used to determine that the lens was operating in the stable regime. 

\begin{figure}
  \begin{center}
    \includegraphics[width=0.8\textwidth]{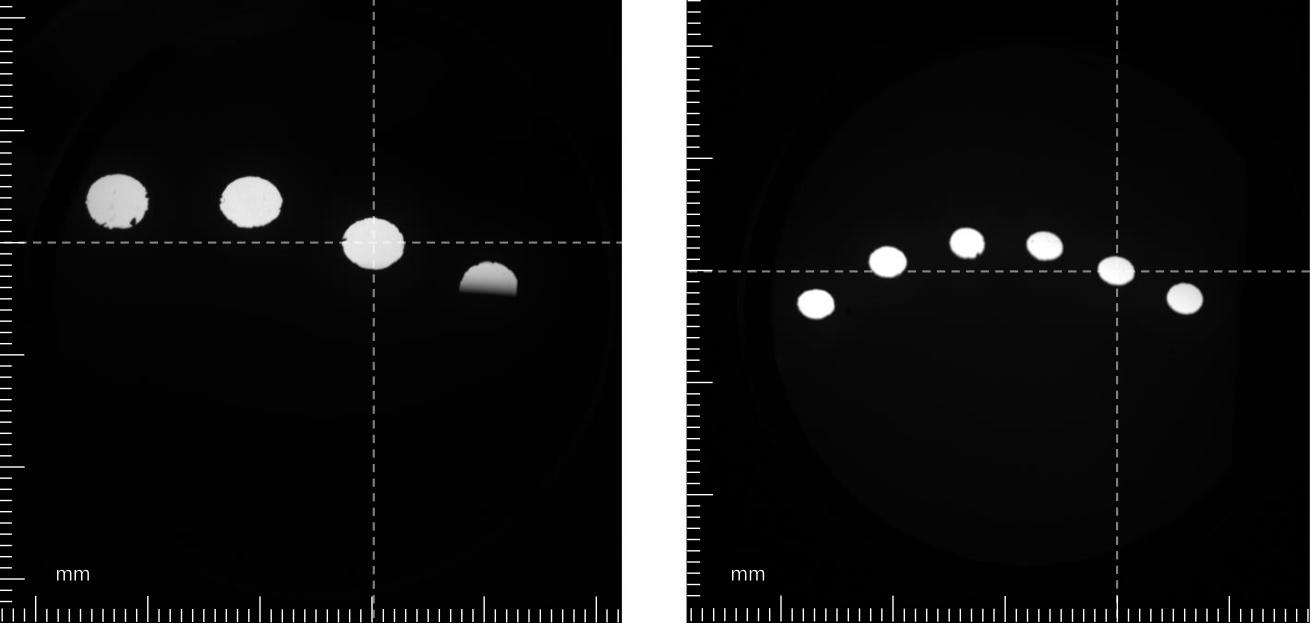}
  \end{center}
  \caption{
    Observed camera image of the 3 beam spots beyond the aperture in
    the Setup 1 configuration, left, and with 6 beam spots beyond the
    aperture in the Setup 2 configuration, right.
    Both images were taken with the lens off. The dashed lines indicate the beam axis and the central beam spot.
  }
  \label{3and6Spot}
\end{figure}

Images of beam impinging on the phosphor screen were taken with the
lens turned off in both the Setup~1 and Setup~2 configurations (see
figure~\ref{3and6Spot}).
Distinct ``spots'' are visible that correspond to the beamlets
produced by the holes in the aperture plate.
The longer drift introduced in the Setup~2 configuration results in a
larger number of beamlets being observed at lower magnification than
in the Setup~1 configuration.
The central axis of the lens passes through the centre of the second
beamlet from the right in both Setup~1 and Setup~2.
Measurement of the diameter of the beam spots and the centre-to-centre
distances allowed the divergence of the beam to be determined.
The divergence in the $x$ and $y$ directions was determined to be
$x^\prime = 1.6$\,mrad and $y^\prime = 0.5$\,mrad, respectively.

\begin{figure}
  \begin{center}
    \includegraphics[width=0.8\textwidth]{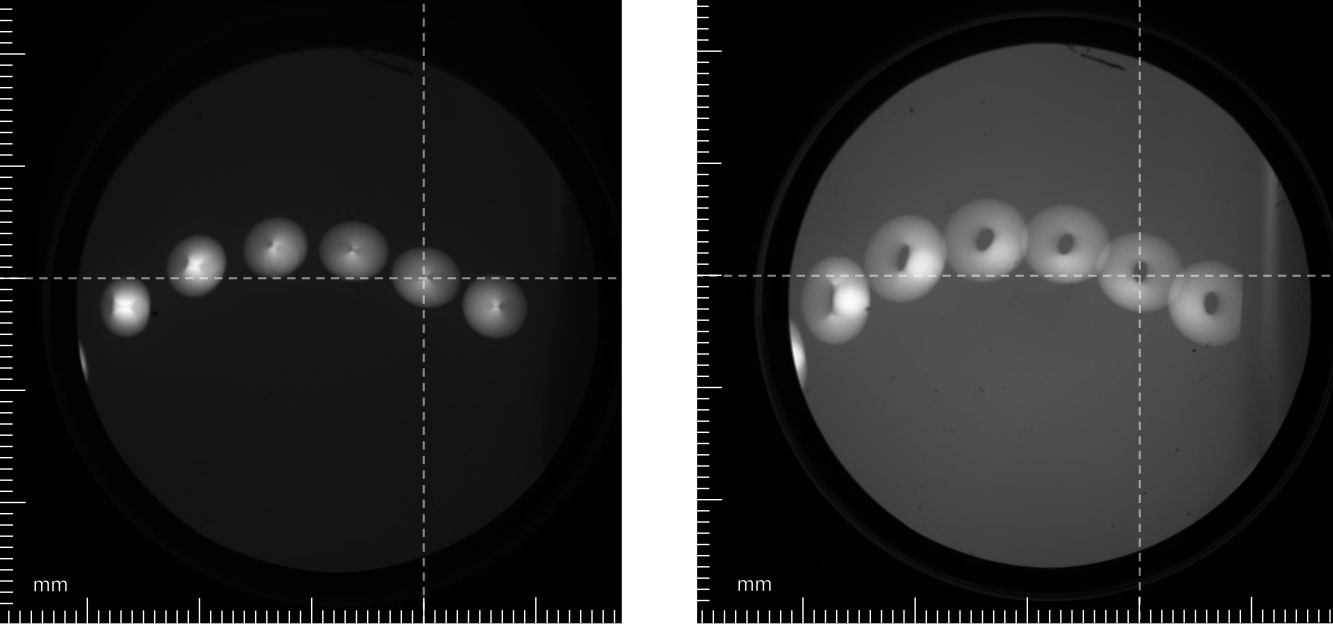}
  \end{center}
  \caption{
    Observed camera image of the 6 beam spots beyond the aperture in
    the Setup 2 configuration with the lens on at a current through
    the coils of 28\,A, left and 33\,A, right.
    Both images were taken with a lens voltage of 20\,kV.
    An additional spot is visible on the left hand side, as the lens
    focusing is increased. The dashed lines indicate the beam axis.
  }
  \label{6SpotLensOn}
\end{figure}

\begin{figure}
  \begin{center}
    \includegraphics[width=0.6\textwidth]{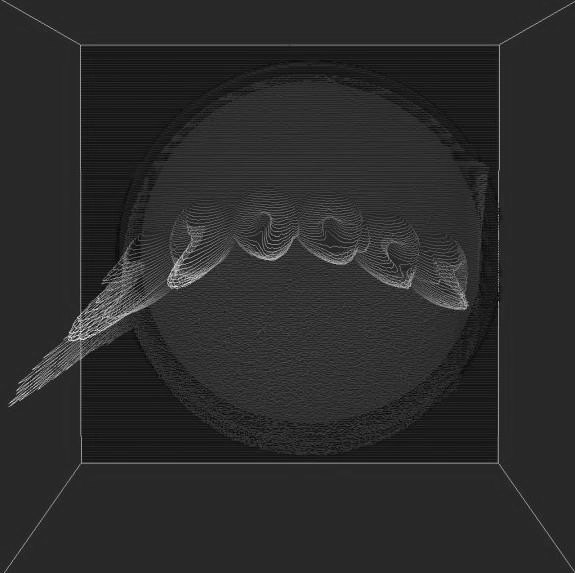}
  \end{center}
  \caption{
    3D plot of the scintillator measurement of the 6 beam spots in the
    Setup 2 configuration, with the lens on.
    The image is shown looking down along the beam axis.
  }
  \label{3DGabor}
\end{figure}

Images of the beam with the lens operating at a voltage of 20\,kV are
shown for currents of 28\,A and 33\,A in figure~\ref{6SpotLensOn}.
The figure shows that the effect of the lens is to produce ring-like
structures on the phosphor screen.
The diameter and eccentricity of the rings increases with radial
distance from the beam axis.
The brightness of the image is observed to vary around the ring.
This effect is seen more clearly in figure~\ref{3DGabor} which shows
the intensity distribution plotted as a function of position on the
phosphor screen.
The ring-like structure of the beam spots is clearly visible against
the low background and the non-uniformity of the intensity
distribution is also observed.
The integrated intensity as well as the intensity distribution around
the ring differs from ring to ring. Similar behaviour has previously been reported
in~\cite{PhysRevLett.85.4518,Posocco:2016uea}.

\begin{figure}
  \begin{center}
    \includegraphics[width=1\textwidth]{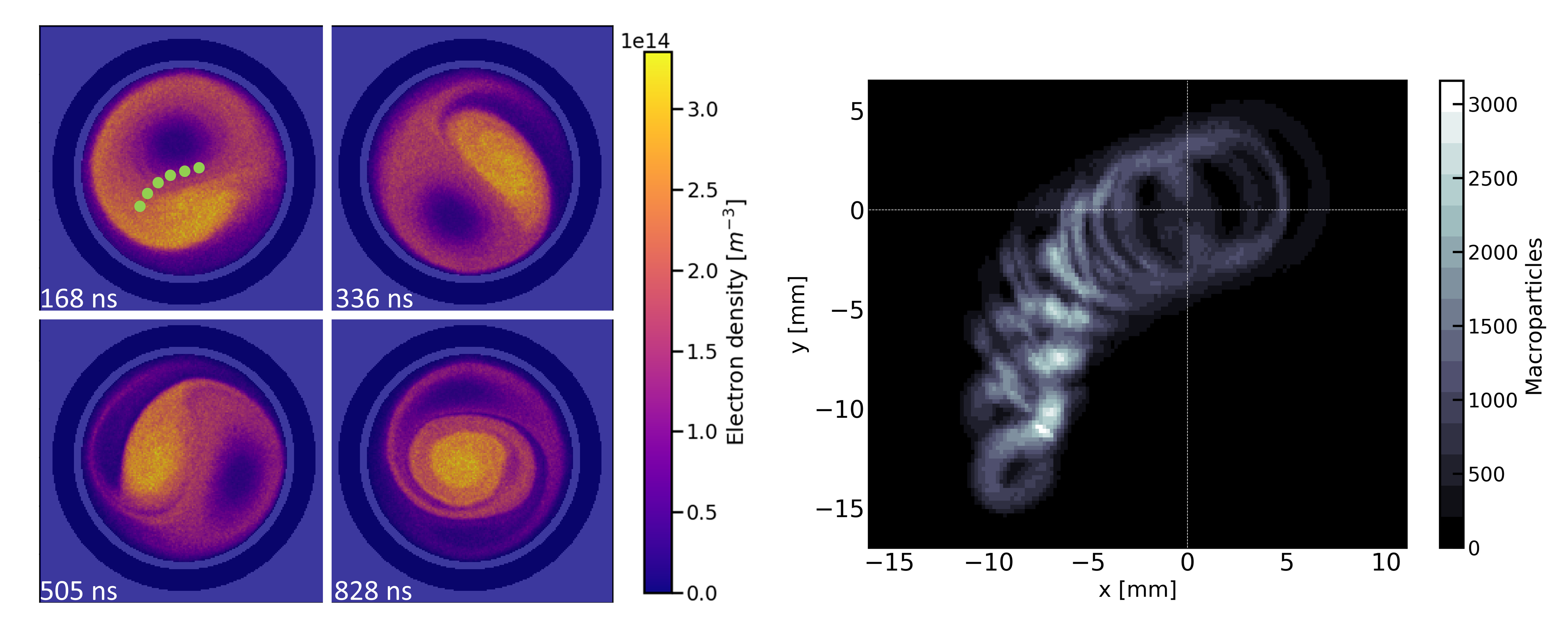}
  \end{center}
  \caption{PIC~\cite{VSIM} simulation of a plasma instability that was observed to focus the proton pencil beams into rings. Left: The averaged density of plasma in a transverse cross-section through the lens at four different time steps during the evolution of the instability. The green spots mark the entry position of the pencil beams. Right: Number of macroparticles hitting a screen 67\,cm downstream of the lens.}
  \label{fig:Dipole_instability}
\end{figure}

To understand better the distribution of the space charge inside the lens and the plasma dynamics during the beam test, a particle-in-cell (PIC) code VSIM~\cite{VSIM} was used to study the main characteristics of a plasma instability that converts pencil beams into rings. The geometry of the central anode, the two end electrodes, and the vacuum tube was reproduced in VSim. A 3D magnetic field map obtained using the actual configuration of the coil was imported from a separate finite-element-analysis package. Thus, the field map described the radially confining magnetic field. The voltage on the central anode was set such that the longitudinal-confinement condition imposed a limit on the maximum electron density which was equal to that imposed by the radial-confinement theory.

The electron cloud was modelled~\cite{CCAP-TN-ACCL-05} as a collision-less plasma using the particle-in-cell (PIC) method in VSim~\cite{VSIM}. The electrostatic potential was calculated from the charge density by solving Poisson's equation on a 3D grid. Then, the macro-particles were advanced in time according to the Lorentz equation. The time step was set to $0.2\,\mathrm{ns}$ (corresponding to $\approx 6 \tau_c$, where $\tau_c = 2\pi m/eB$ is the cyclotron period) to track the electron movement correctly. The grid had a transverse cell size $\Delta x=0.14\lambda_D$ and longitudinal cell size $\Delta z=0.8\lambda_D$, where $\lambda_D$ is the Debye length. Each macro-particle represented approximately 2000 real electrons.

To drive an instability the electrons were loaded at the beginning of the simulation as a plasma column displaced from the central axis or with a large positive radial gradient in the electron density. The six, 1.4\, MeV, proton pencil beams were modelled as macro-particles entering the lens and left to propagate through the electron plasma. The simulation registered the distribution of the proton macro-particles that hit the exit plane of the lens. These macro-particles were then tracked separately through an additional drift space of 67\,cm.

The protons were propagated through the electron plasma using VSIM to simulate the impact of a number of plasma instabilities that have been observed experimentally~\cite{Kapetanakos, Rosenthal}: a hollow electron ring and the diocotron instability~\cite{Levy1965}. The diocotron modes observed in the simulations corresponded to higher order modes with an azimuthal mode number $\mathrm{m}> 1$. Within the range of electron densities between $1\times10^{13}$ m$^{-3}$ and $1\times10^{15}$ m$^{-3}$, no ring formation was observed in the simulations. The instabilities named above show good azimuthal symmetry during their evolution and, hence, would focus the pencil beams to the same position at all times. A displacement of the bulk of the plasma from the central axis and the rotation of the focusing centre are necessary for the formation of rings. An example of such an instability is shown in figure~\ref{fig:Dipole_instability} and consists of a region of high electron density and a region of low electron density that rotate around the beam axis. Figure~\ref{fig:Dipole_instability} shows the result of tracking six proton pencil beams through the instability. Rings are formed on a screen downstream of the lens. In the PIC simulation, the instability was gradually damped due to the absence of a driving mechanism. As the bulk of the plasma approaches the central axis, each pencil beam is focused on a ring with a radius that decreases and a centre that shifts with time. Thus, in the simulation, each pencil beam is transformed into a set of overlapped rings. By contrast, in the experiment, each pencil beam produced a single ring. This experimental outcome was observed consistently throughout the two days of the beam test.

 A Monte Carlo particle-tracking code, BDSIM~\cite{bdsim}, was used to simulate the formation of the rings on a screen downstream of the lens for a simplified plasma distribution. The electron cloud was modelled according to the main features of the $\mathrm{m}=1$ diocotron mode, as a longitudinal column of electrons with azimuthal symmetry. The central axis of the column was displaced from the beam axis and rotated around the beam axis with a constant period that was larger than the transit time of the protons through the lens. The electron cloud had a region of constant density in the centre and a negative radial gradient up to the walls of the anode~\cite{CCAP-TN-ACCL-05}. A 3D time-dependent electric field map was calculated from the rotating plasma column. To study the focusing effect of the plasma and the lens, the six proton pencil beams were tracked through the electric and magnetic field maps. The initial phase-space of the protons upstream of the lens was tuned such that the intensity profile of the pencil beams on the screen obtained from the simulation matches the images taken during the beam test with the lens off.

\begin{figure}
  \begin{center}
    \includegraphics[width=1.0\textwidth]{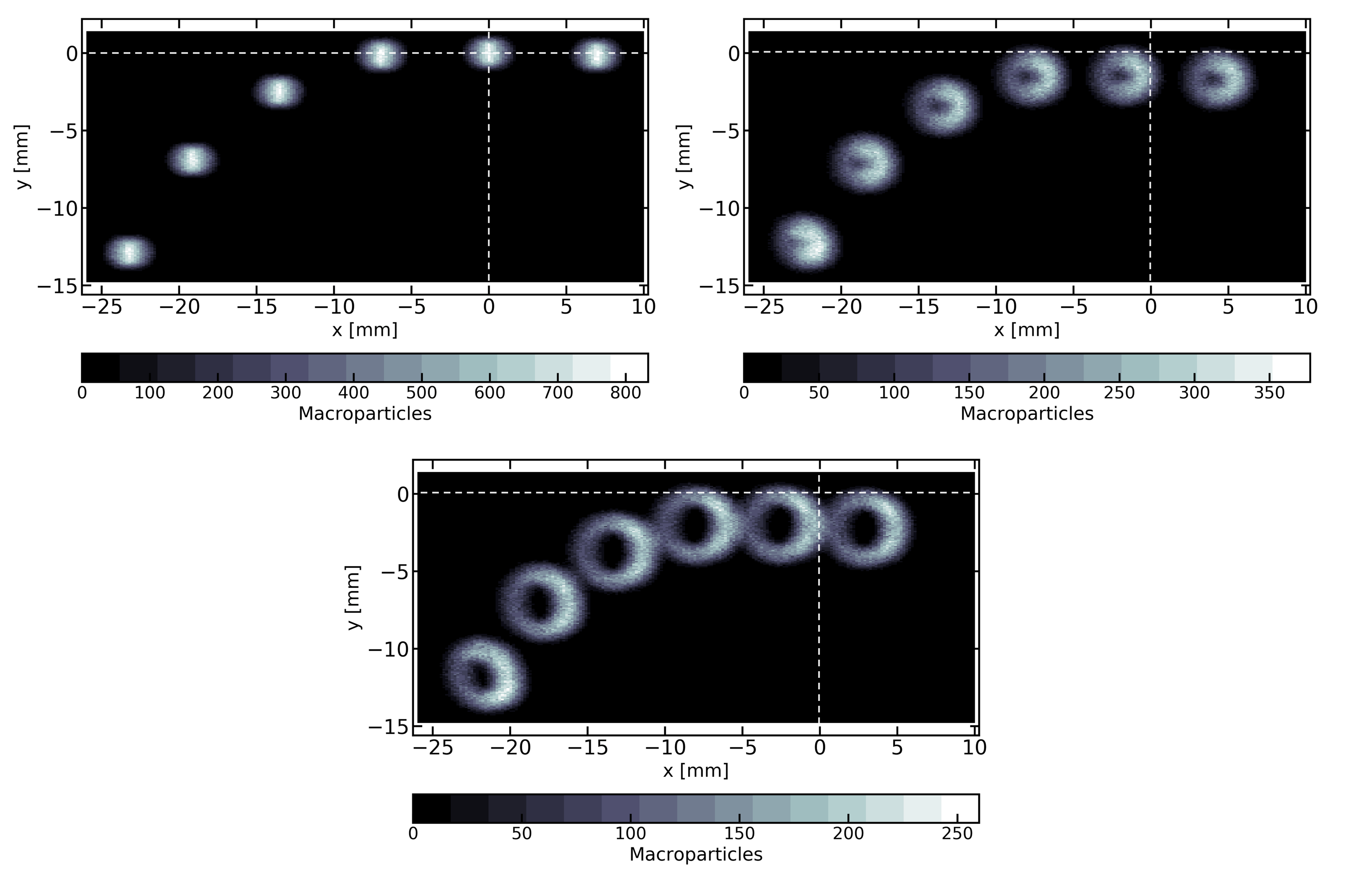}
  \end{center}
      \caption{The effect of a plasma column rotating around the beam axis on six proton pencil beams as simulated with BDSIM~\cite{bdsim} for electron densities of $0$\,m$^{-3}$, $1.8\times10^{14}$\,m$^{-3}$, and $2.8\times10^{14}$\,m$^{-3}$. Increasing the density of the plasma modifies the separation and the width of the rings. The plasma column has a radius $\mathrm{r_c} = 14 $\,mm and an offset $\mathrm{D} = 7$\,mm from the central axis of the lens.} 
  \label{fig:Rings_sim}
\end{figure}

Figure \ref{fig:Rings_sim} shows the focusing effect of a rotating plasma column on the pencil beams as a result of the particle tracking. A variation of the separation between the rings and the width of each ring is seen as a function of the density of the plasma. The shape and thickness of the rings are influenced by changes in the electron density and in the radius of rotation of the plasma column. The eccentricity of the rings increases for the pencil beams that are further away from the beam axis as a result of the different focusing strengths in the $x$ and $y$ directions. This geometrical effect depends on the relative position of the pencil beam with respect to the rotation axis of the plasma column. As in the experimental observations, the brightness of each ring is seen to vary along the circumference. The simulations indicated that the position and extent of the intensity peak is dependent on the ratio between the period of rotation of the plasma column and the transit time of the protons through the lens.

A systematic study of the characteristics of the rings was carried out
as a function of the voltage and current settings at which the lens
was operated during the beam test.
Three parameters were used to characterise the rings:
\begin{enumerate}
  \item Centroid $\mathbf{r_c} = \left( x_c, y_c \right)$:
    The centroid was taken to be the weighted average of all the pixels constituting a ring above a fixed intensity threshold $ (x_c, y_c) = \left( \frac{M_{10}}{M_{00}},\frac{M_{01}}{M_{00}} \right)$, where $M_{ij}$ are the image moments $M_{ij} =  \sum_{x} \sum_{y}  x^i y^j I(x,y)$ of the pixel intensity $I(x,y)$.
  \item Diameter, $\mathcal{D}_{x,y}$: The diameter of the ring (or of
    the beam spot in images taken with the lens off) was determined
    along the $x$ and $y$ directions separately. The diameter is defined as the width of a beam spot or ring along the $x$ or $y$ direction after an intensity cutoff was applied to a camera image. 
  \item Eccentricity, $\mathcal{E}$: The eccentricity is defined as
    the ratio $\frac{\mathcal{D}_x}{\mathcal{D}_y}$.
\end{enumerate} 
In order to extract the diameter and the position of the centroid of each ring, an intensity cutoff was applied to the camera images. This procedure introduces an associated uncertainty in the calculations. For the parameters given above, the uncertainties are $\pm 0.3$\,mm for $\mathcal{D}_{x}$, $\pm 0.2$\,mm for $\mathcal{D}_{y}$, and $\pm 0.05$\,mm for $x_c$ and $y_c$.

\begin{figure}
  \begin{center}
    \includegraphics[width=0.8\textwidth]{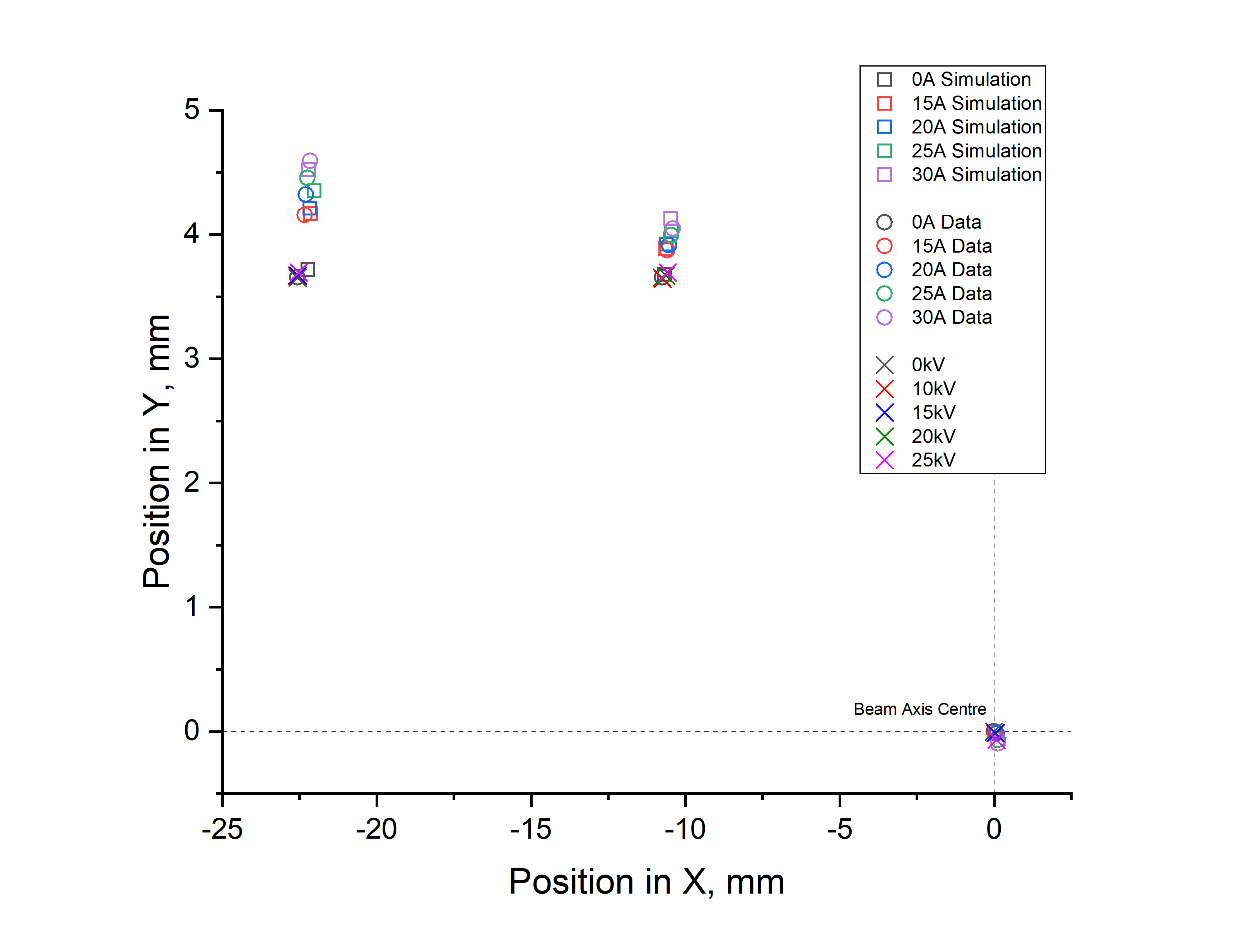}
  \end{center}
  \caption{
    Position of the centroid of the 3 beam spots for varying magnetic
    fields and high voltages. Squares and Circles represent variation in current through the
magnetic coils only, in simulation and data respectively, with no
applied high voltage. Crosses represent variation in high voltage (and therefore electric
field) with no current through the magnetic coils.
  }
  \label{fig:GaborPlot}
\end{figure}

A comparison of the effects of electric field only and magnetic field
only in the lens is shown in figure~\ref{fig:GaborPlot}.
This plot presents data from Setup 1, in which the 3 spots are those
shown in figure~\ref{3and6Spot}, with the rightmost point
corresponding to the beam axis. As expected, applying only an electric
potential does not influence the particle transport, while a variation in the magnetic field leads to a more significant change in the focusing strength.
Comparison between magnetic field only data and the results of particle transport in the magnetic field shows good agreement in
direction as well as in magnitude.
This is true for all three of the observed pencil beamlets.
The small variation around the pencil beam on the beam axis indicates
that the beam axis, aperture axis, and lens axis were not identical. 

\begin{figure}
  \begin{center}
    \includegraphics[width=0.8\textwidth]{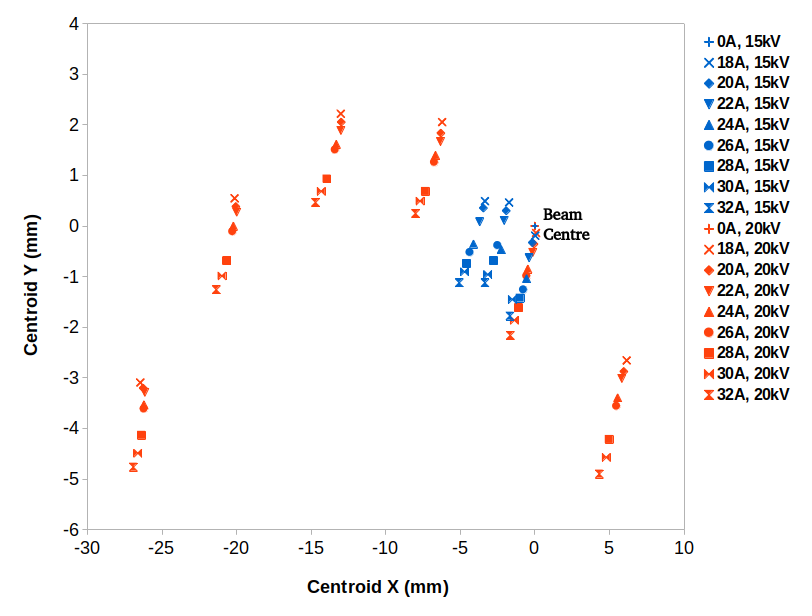}
  \end{center}
  \caption{
    The $x$ and $y$ position of the centroids of the beam spots imaged with the Setup 1 (blue) and Setup 2 (red) with increasing magnetic field strength.
  }
  \label{fig:CentroidXVS}
\end{figure}

Figure~\ref{fig:CentroidXVS} shows the variation in $x$ and $y$ position
of the three beam spot centroids in Setup 1, under the effects of both
applied magnetic field and high voltage.
In this case, the high voltage is held at 15\,kV, while the current
through the magnetic coils is varied from 0 to 32\,A. 
An approximately linear increase with magnetic field is observed. In addition to the circular motion of the plasma in the lens, there is a focusing force that increases with the external magnetic field. A possible cause for this effect is a misalignment between the beam axis and the symmetry axis of the coil. The two sets of data, for 15\,kV and 20\,kV, were taken on consecutive days which indicates that the driving mechanism associated with the observed plasma dynamics is a characteristic of the geometry and operation of the lens. 

\begin{figure}
  \begin{center}
    \includegraphics[width=0.8\textwidth]{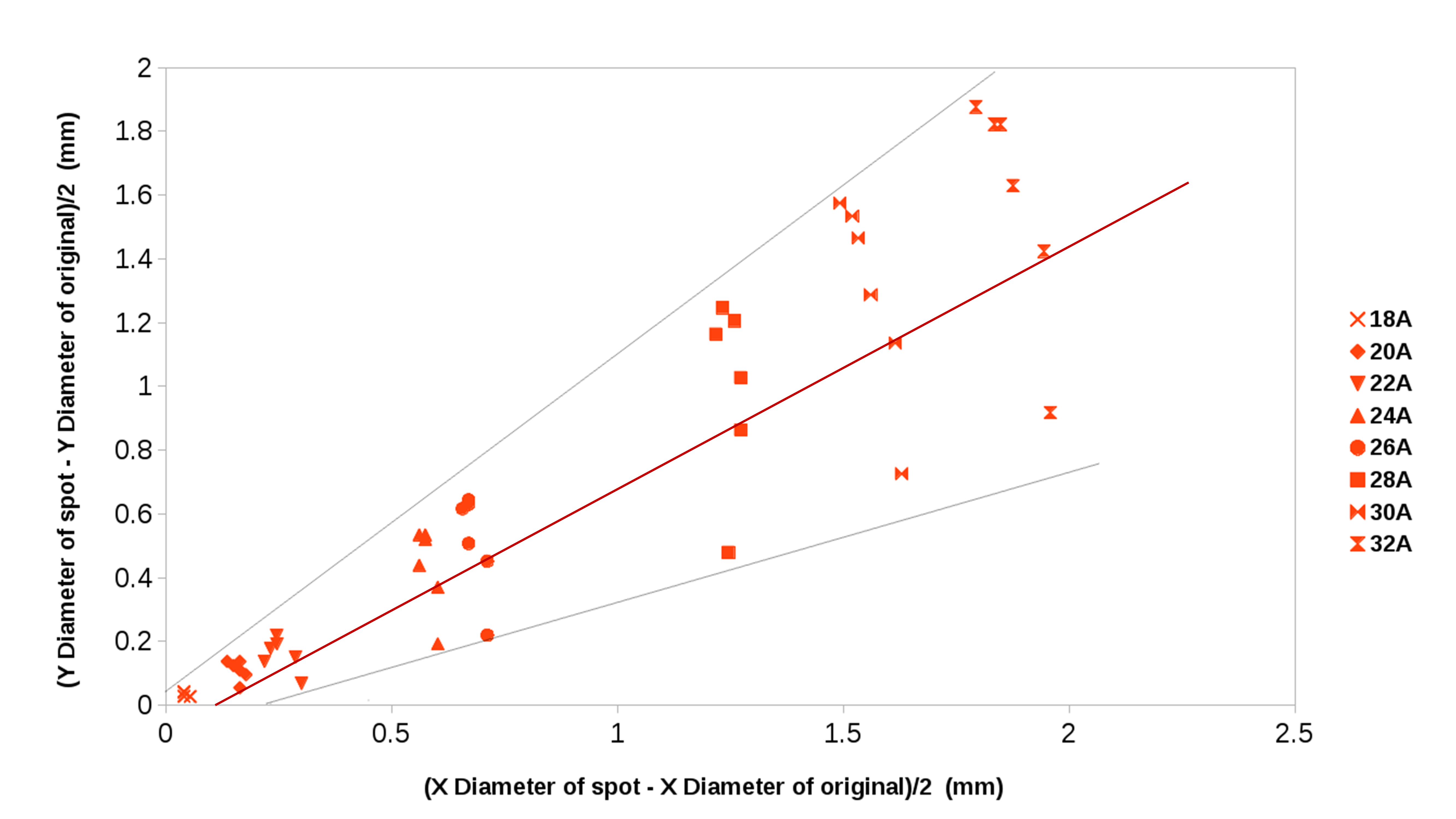}
  \end{center}
  \caption{
    The variation in the $x$ and $y$ diameter of the 6 spot data with
    increasing magnetic field strength.
  }
  \label{fig:DiameterDifferences}
\end{figure}

The variation in $x$ and $y$ diameter of the six beam spots from Setup 2
is given in Figure~\ref{fig:DiameterDifferences}: with a constant
applied voltage of 15\,kV, the variation in spot diameter with magnetic
field is shown.
A non-linear increase in spot size with change in magnetic fields is
observed, with the rate of increase in diameter getting larger at high
magnetic fields. This indicates that the increase in plasma density with magnetic field is faster than linear and thus that the plasma trapping efficiency varies with the magnetic field strength.
Since the points remain within the lines shown, there is an indication of a trend for the change in diameter for a given spot.
The ratio variation being solely dependent on initial position of the spot indicates that this effect is caused by the density
distribution of the plasma in the transverse plane. 

\begin{figure}
  \begin{center}
    \includegraphics[width=0.8\textwidth]{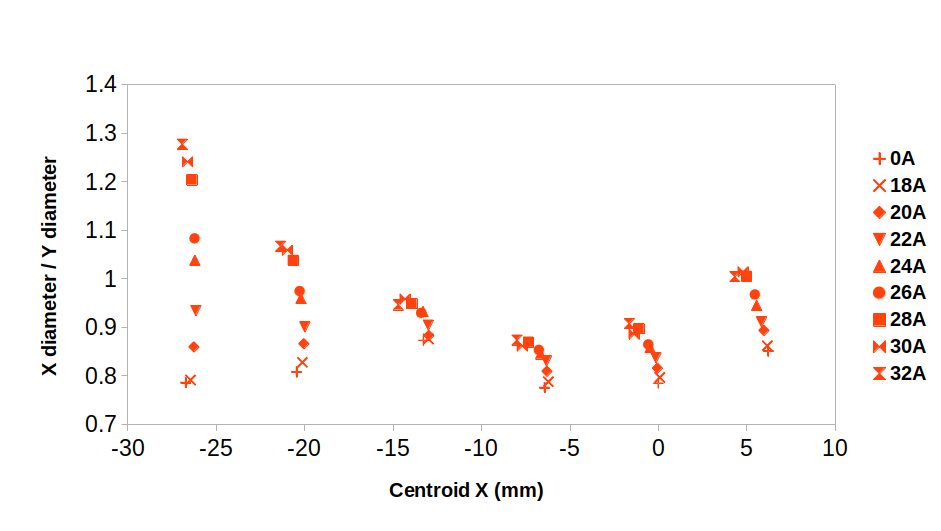}
  \end{center}
  \caption{
    The ratio of the $x$ diameter to the $y$ diameter for beam spots in
    the 6 spot data with increasing magnetic field strength.
  }
  \label{fig:XYRatio}
\end{figure}

Figure~\ref{fig:XYRatio} shows the change in $xy$ ratio of the six beam
spots from Setup 2 with variation in magnetic field.
The focusing force in the $x$ direction compared to the focusing force in the $y$ direction changes differently for each beam spot. The difference between the two focusing forces is more significant for those beam spots at greater distances from the beam axis. This indicates that the centre of the lens has a low plasma density, while further from the axis of the lens, the plasma density increases with radius.
In addition, there is some perturbation that causes the motion of the plasma to evolve with time. Figures~\ref{fig:GaborPlot} to~\ref{fig:XYRatio} may be understood in terms of a position dependent plasma density, with the bulk of the plasma shifted radially from the lens centre and rotating around the central axis.

\section{Conclusions}
We characterised the operation of an electron plasma (Gabor) lens based on a test at the Surrey Proton Beam Centre with a 1.4\,MeV proton beam. Prior measurements at Imperial College indicated that the lens had a stable regime of operation over a range of applied voltage and current. During the beam test, the lens was observed to transform pencil beams into rings. The presence of a plasma was confirmed by matching the measured focusing effect with particle transport calculations. An evaluation of the focusing strength showed that the density of the trapped electrons depends on the strength of the radially confining magnetic field, effects that were well described by simulation. Since the same focusing effects and ring patterns were observed on consecutive days, the plasma instability was associated with the geometry and operation of the lens.

The formation of rings indicates that the plasma column is excited into a coherent off-axis rotation. The size of the rings increases with increasing current through the coil. A reproducible modulation of the intensity was observed around the circumference of each ring. The position of the centroids of the rings varied non-linearly with the external magnetic field strength, showing a variable plasma trapping efficiency. The $x$ and $y$ diameters, and the eccentricity of the rings were seen to depend on their position with respect to the beam axis, as a result of the different $x$ and $y$ focusing forces experienced by each pencil beam.

Both particle-in-cell and particle-tracking simulations showed that a rotation of the bulk of the plasma transforms pencil beams into rings. The size and width of the rings were shown to be determined by the density of the plasma. Rings with size, eccentricity, and intensity modulation similar to the experimental images were reproduced with a simulation of particle transport through a plasma characterised by the $m=1$ diocotron instability.

The results described here indicate the presence of a mechanism that drives the rotation of the plasma column. Further investigations are required to identify and describe the exact mechanism that needs to be avoided for the lens to be operated as a reliable focusing device.
\section*{Acknowledgements}

We wish to acknowledge the work of Piero Posocco at the 2017 beam test. 
We would also like to the thank all the staff at the University of Surrey Ion Beam Facility, and in particular Vladimir Palitsin, for his support throughout both days of the beam test. We gratefully acknowledge the Plasma Physics Group at
Imperial College London who were generous in sharing 
their scientific insight and provided technical support
at various stages of the project. The work was financially supported by the STFC through the Imperial Impact Acceleration Account. One of us (TSD) gratefully acknowledges the support from the Class of 1964 Scholarship.

\bibliographystyle{99-Styles/utphys}
\bibliography{GaborPaper}

\end{document}